# Superconductivity in $WO_{2.6}F_{0.4}$ synthesized by reaction of $WO_3$ with Teflon

D. Hirai, E. Climent-Pascual and R. J. Cava
Department of Chemistry, Princeton University, Princeton, NJ 08544, USA

## Abstract

$WO_{3-x}F_x$ ($x \leq 0.45$) perovskite-like oxyfluorides were prepared by a chemically reducing fluorination route using the polymer polytetrafluoroethylene (Teflon). The symmetry of the crystal structures of $WO_{3-x}F_x$ changes from monoclinic to tetragonal to cubic as the fluorine content increases. Fluorine doping changes insulating $WO_3$ to a metallic conductor, and superconductivity ($T_c$ = 0.4 K) was discovered in the samples with fluorine contents of $0.41 \leq x \leq 0.45$. This easy fluorination method may be applicable to other systems and presents an opportunity for finding new oxyfluoride superconductors.

# Introduction

Transition metal oxides exhibit a wide range of electronic and magnetic ordering behavior at low temperatures, which is usually highly dependent on carrier concentrations. Chemical doping is the most frequently used technique to control carrier concentrations, and hence the electronic state obtained. Fluorine has one additional electron compared to oxygen and, due to the compatible sizes of $F^-$ and $O^{2-}$, may be considered a universal dopant for oxides. Due to the difficulties encountered in oxyfluoride synthesis, however, it has not been commonly employed as such. The prototype iron based high temperature superconductor LaFeAs(O,F) is an excellent example of using fluorine to alter a mother compound (originally magnetic) to induce a new electronic state [1].

In the past, a variety of fluorination techniques have been invented to dope fluorine or synthesize oxyfluorides. Oxidative fluorination through intercalation of $F^-$ obtained by decomposition of unstable fluorides such as $CuF_2$ or $ZnF_2$ has been employed to synthesize oxyfluoride cuprate superconductors, for example [2]. High temperature synthesis using binary fluoride starting materials such as $MF_2$ or $MF_3$ (M = Alkali-earth or lanthanide metal) is a common synthetic route for obtaining oxyfluorides [1,3], but this method is limited due to the high stability of the starting binary fluorides relative to the intended products. In order to overcome this problem, high pressure synthesis or low temperature fluorination has been developed. High pressure synthesis has been successful for example in the preparation of superconducting cupper oxyfluorides [4,5], but has the disadvantages of requiring special equipment and generating relatively small sample amounts. Reactions of oxides with aqueous and anhydrous hydrofluoric acid (HF) at ambient or at high pressure have also been employed, with the disadvantages of handling HF in a laboratory environment.

Here we employ a recently described fluorination route that uses a powder of the organic polymer polytetrafluoroethylene (PTFE) (Teflon) as a source of fluorine [6] in a reductive fluorination process. Several important characteristics of PTFE, i.e. its solid form and chemical stability at room temperature and its relatively low decomposition temperature compared with inorganic compounds, enable us to perform reductive fluorination of oxides easily and safely. In order to demonstrate the convenience and usability of this method for potentially superconducting electronic materials, the simple binary transition metal oxide tungsten trioxide was selected as a basis for this study. The systematic preparation of $WO_{3-x}F_x$ has already been reported [7]. However, $WO_{3-x}F_x$ was obtained in that study by heating $WO_3$ with HF under the extreme condition of 3000 atmospheres in a sealed gold ampoule. Although $WO_{3-x}F_x$ was synthesized with various fluorine contents, the physical properties, especially at low temperatures, were not investigated completely. In the current study, $WO_{3-x}F_x$ was successfully synthesized through reaction with PTFE. We find that as the fluorine content increases in $WO_{3-x}F_x$, the crystal structure changes from monoclinic, to tetragonal, to cubic, while simultaneously the conductivity changes from insulating to metallic. The tetragonal symmetry phase has not been previously reported. Furthermore, a superconducting transition was discovered below 0.4 K in cubic $WO_{3-x}F_x$ with high fluorine content. The discovery of superconductivity in this fluorine-doped simple oxide demonstrates that the PTFE fluorination technique may provide a unique route for finding new superconductors.

**Experimental**

A pellet of tungsten trioxide (99.8 %; Alfa Aesar) was placed in an evacuated quartz tube with an excess amount of PTFE powder and sintered at 550 °C for 36 hours. The reaction was performed on approximately 100 mg of tungsten trioxide in a quartz tube with a volume of 15 cm$^3$. The fluorine and tungsten contents of representative samples were checked after completion of the reaction by pyrohydrolysis and ICP-OEA. These chemical analyses were performed by Galbraith Co. The purity, symmetry, and cell parameters of the samples were evaluated by means of x-ray powder diffraction data collected at room temperature on a Bruker D8 FOCUS diffractometer (Cu K$\alpha$) over a $2\theta$ range between 5 and 110° with a step size of 0.014°. Lattice parameters, atomic positions, and atomic displacement factors were refined by the Rietveld method using the FULLPROF program integrated in the WINPLOTR software [8]. Oxygen and fluorine were assumed to be distributed randomly, and their fractions were fixed at the values obtained from chemical analysis. Diffraction maxima were fit with the Thompson-Cox-Hastings pseudo-Voigt function starting from the instrumental resolution values for the profile parameters U, V and W. These starting profile parameters were obtained by fitting data obtained for sintered samples of an $Al_2O_3$ standard. The background was characterized by use of a 6-coefficient polynomial function. Hall measurements were performed in a physical property measurement system (PPMS: Quantum Design). Resistivity and specific heat measurements down to 0.35 K were performed in a PPMS equipped with a $^3$He insert.

Results and Discussion

The reaction with $WO_3$ and polytetrafluoroethylene (PTFE) occurs at the gas-solid interface. PTFE begins to decompose slowly at 260 °C, and more rapidly above 400 °C [9]. The main product of decomposition under vacuum around the reaction temperature has been reported to be tetrafluoroethylene (TFE) monomer [10]. TFE produced by the decomposition of PTFE evaporates and fills the quartz tube at 550 °C. However, the actual fluorination mechanism is unclear because secondary reactions of TFE generate several kinds of fluorocarbons, such as hexafluoropropene and cyclo-perfluorobutane, depending on the temperature and reaction pressure [10].

The fluorine content of several samples that were found to be single phase by x-ray measurement was determined by chemical analysis. The nominal and actual tungsten to fluorine ratios are listed in Table 1. With increasing nominal fluorine content, the actual fluorine content $x$ in $WO_{3-x}F_x$ increases. However, the relationship between the nominal content and resulting composition is not linear, and a large excess of PTFE is required to achieve high fluorine content in $WO_{3-x}F_x$. This may be because thermodynamically stable fluorocarbons are formed under PTFE-rich conditions. Excess amounts of PTFE (likely at least partially decomposed into a variety of fluorocarbons) covered the interior of the quartz tubes after the reactions.

The maximum fluorine content achieved by this method was $x = 0.45$ in $WO_{3-x}F_x$. Attempts to achieve higher fluorine content by increasing the amount of PTFE or the sintering temperature resulted in over-reduction and the appearance of $WO_2$ as an impurity phase. This maximum fluorine content is smaller than that achieved in previously reported reaction methods, for example with HF under high pressure, which can prepare fluorine content up to $x = 0.43$ in 48 % aqueous HF, and $x = 0.66$ in anhydrous HF [7]. The Teflon reaction method is notably easier and safer than performing reactions with HF.

As shown in the XRD patterns for different fluorine contents (Figs. 1(a) and (b)), the crystal structure of $WO_{3-x}F_x$ depends highly on the fluorine content $x$. The XRD patterns of $WO_{3-x}F_x$ for $x = 0.11$ and $x = 0.41$ can be well indexed on the tetragonal (space group: $P4/nmm$) and cubic (space group: $Pm\bar{3}m$) unit cells, respectively. The crystallographic information and Rietveld agreement factors for $WO_{3-x}F_x$ for $x = 0.11$ and $x = 0.41$ are summarized in Table 2. The cubic $WO_{3-x}F_x$ phase in the range of high fluorine content ($0.17 \leq x \leq 0.66$) has been reported previously [7], but the tetragonal phase between monoclinic $WO_3$ and cubic $WO_{3-x}F_x$ has not previously been reported. Although the agreement between the calculated and observed XRD patterns is reasonable, it is difficult to determine the positions of the light elements oxygen and fluorine accurately from the XRD data. Detailed structural studies using electron or neutron diffraction for related compounds tetragonal $Na_{0.1}WO_3$ [12] and cubic $Li_{0.36}WO_3$ [13] have revealed that small tilting or rotation of $WO_6$ octahedra results in a larger unit cell than that determined by XRD measurement [14]. Neutron diffraction data would be required to more accurately determine the oxygen/fluorine positions for the current phases.

The crystal structure of $WO_{3-x}F_x$ becomes successively higher in symmetry as the fluorine content increases. The relation of the monoclinic, tetragonal and cubic cells of $WO_3$, $WO_{2.89}F_{0.11}$ and $WO_{2.59}F_{0.41}$ is schematically illustrated in Fig. 1(d). The relationships are: $a_M, b_M, c_M \sim \sqrt{2}a_T \sim 2c_T \sim 2a_C$. The crystal structure of $WO_3$ is formed by $WO_6$ octahedra, and each $WO_6$ octahedron is part of a corner-sharing network extending in three dimensions. The distortion of each octahedron results in monoclinic symmetry, with 8 formula units in each unit cell. As the distortion of the $WO_6$ octahedra becomes more regular with F doping, the unit cell becomes tetragonal with a small deformation of the octahedra due to different W-O bond lengths along the $c$-axis. Finally a simple $ReO_3$-type cubic cell with one W atom and 180 degree W-O-W bonds (Fig. 1(c)) is formed. It is informative to compare cell volumes per single formula unit, because all the crystal structures are closely related. Lattice constants and cell volumes divided by $Z$ are listed in Table 1. The normalized cell volume shows a monotonic increase from monoclinic to tetragonal to cubic cell. Although the ionic radius of $F^-$ (1.29 Å) is slightly smaller than that of $O^{2-}$ (1.35 Å) [15], the reduction of $W^{6+}$ during fluorination to different fractions of (formally) $W^{5+}$ with a larger radius leads to an expansion of cell volume.

The lattice constant also increases with increasing F content in the cubic phase; this can be used as a good indicator to compare fluorine contents in that phase. We follow the behavior in the cubic $Na_xWO_3$ ($0.5 < x$

< 1.0) phase, where a linear relationship between lattice constant and sodium content has been established [16]. Assuming an analogous linear relationship between lattice constant and the analytically measured fluorine content in the cubic $WO_{3-x}F_x$ phase, a line based on the data for $x =$ 0.41 and x = 0.45 ($a = 3.792 + 0.045 \times x$ (Å)) enables us to roughly estimate the fluorine content from the measured lattice constants. In this paper, the fluorine content in the cubic phase regime for samples whose fluorine content was not measured directly by chemical analysis was estimated by using the formula above.

The light green colored starting $WO_3$ powder turned dark green in the fluorinated tetragonal phase and deep blue in the fluorinated cubic phase. This systematic color change is the same as that generally observed in tungsten bronzes and indicates that charge carriers are introduced into the system. The change in physical properties with fluorine doping can clearly be seen in the resistivity measurements, as shown in Fig. 2(a). The undoped compound $WO_3$ is a band insulator with a band gap of 2.6 eV [17], however, as the fluorination proceeds, the resistivity $\rho(T)$ at 300 K systematically drops, reaching several mΩcm at the highest fluorine content in the cubic phase. The value of the resistivity in the cubic phase is in the regime of a metal, considering the fact that the sample is a sintered polycrystalline pellet. The temperature dependence of the resistivity also changes on fluorination. The resistivity increases exponentially in the tetragonal phase on cooling, indicating the existence of an energy gap, whereas the ratio $\rho(300 \text{ K}) /\rho(5 \text{ K})$ is smaller than 4 in the cubic phase. Even though the magnitude of the room temperature resistivity decreases at high fluorine content, a weak increase of resistivity with decreasing temperature was observed in all samples with a cubic unit cell. This non-metallic behavior may stem from the presence of remnant insulating organics at the grain boundaries or the usual poor connections between grains in lightly-sintered polycrystalline oxide pellets.

In order to evaluate the effects of fluorine doping quantitatively, Hall measurements were conducted for $WO_{3-x}F_x$ ($x = 0.45$) as shown in Fig. 2(b). The inset of Fig. 2(b) shows the magnetic-field dependence of Hall resistivity $\rho_{xy}$ at different temperatures. In the experiment, $\rho_{xy}$ was taken as $\rho_{xy} = (\rho(+H)-\rho(-H))/2$ at each temperature to eliminate the effects of magnetoresistance and misalignment of the Hall electrodes. The negative Hall resistivity $\rho_{xy}$ for $WO_{3-x}F_x$ indicates that electron carriers have been introduced into WO₃, as expected from the substitution of fluorine for oxygen. The observed electron carriers rule out the possibility that fluorine atoms might occupy an interstitial site and behave as hole-donors. $\rho_{xy}$ shows a linear field dependence at all temperatures and the slope obtained by least-squares fit is used to determine $R_H$. The temperature independent $R_H$ is an indication of the metallic nature of WO$_{3-x}$F$_x$ in the cubic phase. Furthermore, the carrier concentration obtained from the Hall coefficient $n$ = 7.16×10²¹ cm⁻³ is sufficient to be a metal. Assuming that all the substituted fluorine atoms contribute to the creation of carriers, the fluorine content determined by chemical analysis ($x$ = 0.45) and the lattice constant ($a$ = 3.814 Å) obtained from the XRD give rise to an estimation of a carrier concentration 8.11×10²¹ cm⁻³. The good agreement between measured and estimated carrier concentration indicates that essentially all of the doped fluorine atoms contribute to produce electron carriers.

As the fluorine content increases, the band insulator WO₃ evolves into metallic WO$_{3-x}$F$_x$ with a cubic structure and a superconducting transition was discovered around $T_c$ = 0.4 K. The evidence for the superconducting transition of WO$_{3-x}$F$_x$ can be observed in the low temperature resistivity data $\rho(T)$, shown in Fig. 3(a). Below $x$ < 0.4, a superconducting transition is not observed down to lowest temperature of our measurement, 0.35 K. However, at $x$ = 0.40, a drop in the resistivity begins to appear as the temperature is lowered to 0.35 K, suggesting the onset of superconducting transition at that temperature (Fig. 3(a)). For the higher fluorine contents of $x$ = 0.41, 0.44 and 0.45, a very clear drop to a zero resistance state indicative of superconductivity is observed at around 0.4 K. The superconducting transition temperature does not vary for $x$ between 0.41 and 0.45. We could not trace the $T_c$ to higher fluorine content because the maximum $x$ obtained was limited by the synthetic method.

The zero resistance state was suppressed by application of a small magnetic field, and the resistivity returned to the normal state value under higher magnetic fields as shown in Fig. 3(b). The zero-temperature upper critical field $H_{c2}(0)$ = 0.53 T was estimated by the initial slope of the upper critical field $H_{c2}(T)$, using the relationship $H_{c2}(0) = -0.7\, T_c\, dH_{c2}/dT_c$ [18]. The obtained $H_{c2}(0)$ = 0.53 T corresponds to the Ginzburg-Landau coherence length $\xi_0 = [\Phi_0 / 2\pi H_{c2}(0)]^{1/2}$ = 250 Å, where $\Phi_0 = hc/2e$ is the magnetic flux quantum. The value of $H_{c2}(0)$ is relatively high, but does not exceed the weak coupling Pauli paramagnetic field $\mu_0 H_P = 1.84\, T_c$ = 0.74 T

[19].

The bulk nature of superconductivity was confirmed by a large jump of specific heat at $T_c$ = 0.4 K (Fig. 3(c)). This jump was completely suppressed under a magnetic field of 1 T, well above $H_{c2}$. The low temperature normal state specific heat can be approximated as $C = \gamma T + \beta T^3$, where $\gamma T$ represents the normal state electronic contribution and $\beta T^3$ represents the lattice contribution to the specific heat. The fitting for the specific heat of $WO_{3-x}F_x$ ($x$ = 0.44) in the temperature range of 0.5 K ≤ $T$ ≤ 5 K yields the electronic specific coefficient $\gamma$ = 1.59 mJ/(mol K$^2$), and $\beta$ = 0.197 mJ/(mol K$^4$). The value of $\gamma$ obtained is relatively small compared with typical transition metal oxides and comparable to that of cubic tungsten bronze $Na_xWO_3$ (0.90 mJ/(mol K$^2$) ≤ $\gamma$ ≤ 1.89 mJ/(mol K$^2$)) [20] indicating weak electron correlation in this system. The Debye temperature $\Theta_D$ = 340 K calculated from $\beta$ is not so different from $Na_xWO_3$ (255 K ≤ $\Theta_D$ ≤ 375 K) [20]. Further study of thermodynamic properties at lower temperature will yield insight into the size and symmetry of superconducting gap.

The systematic changes in crystal structure and physical properties of the $WO_{3-x}F_x$ system are very similar to those of the tungsten bronzes $A_xWO_3$ ($A$ = Na, Li) [21,22]. Both in $WO_{3-x}F_x$ and $A_xWO_3$, electron carriers are introduced by increasing dopant content $x$. The doped cation in $A_xWO_3$ occupies the cavity in between $WO_6$ octahedra, while the fluorine ion substitutes for oxygen forming $W(O,F)_6$ octahedra. Although the substituted atomic position is quite different, the effects on structure and properties are quite similar. At the low doping level of $x$ these systems are semiconductors and possess symmetries lower than cubic. However, when $x$ is sufficiently high, the materials become cubic and metallic. The cubic phase exists over a wide range of compositions with conductivities generally increasing with increasing $x$.

Similarly, superconductivity is also observed in $Na_xWO_3$. Superconducting transitions (0.7 K ≤ $T_c$ ≤ 3 K) have been reported for $Na_xWO_3$ (0.2 ≤ $x$ ≤ 0.4), but it has a tetragonal bronze structure, not a perovskite structure in the superconducting composition regime [23]. Note that no superconductivity has been observed down to 0.1 K in cubic phase $Na_xWO_3$ [24]. It is worthwhile to compare the superconducting transition temperatures in these systems. The $T_c$ of $WO_{3-x}F_x$ is considerably lower than that of $Na_xWO_3$. The significant chemical difference between these systems is the position of the dopant atoms leading to electronic doping. Carriers introduced by chemical doping fill the empty $t_{2g}$ orbital of tungsten. The conduction band in the tungsten bronze is formed by a π-type interaction between the $t_{2g}$ orbitals of tungsten and the appropriate $p$ orbitals of oxygen [25]. In the case of Na$_x$WO$_3$, the doped cation occupies the cavity between the WO$_6$ octahedra. Thus, the doped Na ion does not give rise to disorder in the lattice supporting the conduction electrons. On the other hand, fluorine doping, where the fluorine ion is substituted in the oxygen ion framework, introduces significant disorder into the conduction band. The disordered character of WO$_{3-x}$F$_x$ can be seen in the very high resistivity at low temperatures despite the high carrier concentration of $n = 7.16 \times 10^{21}$ cm$^{-3}$. Even the previously reported resistivity measurements on a single crystal sample show a small resistivity ratio ($\rho$(290 K) / $\rho$(4.2 K) = 1.1) [7] implying a disordered nature of this system. Although we cannot simply compare superconductors in different crystal structures, we speculate that the intrinsic disordered nature inherent to the substitution site might lead to the one order of magnitude smaller $T_c$ in WO$_{3-x}$F$_x$ than in $A_x$WO$_3$. Determination of the origin of the difference in $T_c$ would be of interest in future work.

In conclusion, we have demonstrated the utility of a new fluorination technique using PTFE to introduce electrons into oxides by making a band insulator, WO$_3$, into a superconductor. The crystal structure of WO$_{3-x}$F$_x$ changes systematically from monoclinic to tetragonal to cubic with increasing fluorine content. The fluorine doping significantly affects the physical properties of WO$_3$, and insulating WO$_3$ becomes semiconducting, metallic, and finally superconducting with increasing fluorine content. We believe this convenient technique of fluorination opens a new route of controlling physical properties and finding new superconductors in oxides.

## Acknowledgement

This work was supported by the AFOSR MURI on superconductivity.

**References**

[1] Y. Kamihara, T. Watanabe, M. Hirano, and H. Hosono; J. Am. Chem. Soc. **130**, 3296 (2008).

[2] P. R. Slater, J. P. Hodges, M. G. Francesconi, P. P. Edwards, C. Greaves, I. Gameson, M. Slaski; Physica C, **253**, 16 (1995).

[3] R. L. Needs, M. T. Weller, U. Schelerb and R. K. Harris; J. Mater. Chem. **6**, 1219 (1996).

[4] T. Kawashima, Y. Matsui and E. Takayama-Muromachi; Physica C, **257**, 313 (1996).

[5] M. Isobe, J. Q. Li, Y. Matsui, F. Izumi, Y. Kanke and E. Takayama-Muromachi; Physica C, **269**, 5 (1996).

[6] Y. Kobayashi, M. Tian, M. Eguchi, and T. E. Mallouk; J. Am. Chem. Soc. **131**, 9849 (2009).

[7] A. W. Sleight; Inorganic chemistry **8**, 1764-1767 (1969)

[8] J. Rodr´ıguez-Carvajal and T. Roisnel, FULLPROF, WWINPLOTR, and accompanying programs, 2008; http://www.ill.eu/sites/ fullprof/index.html.

[9] B. B. Baker and D. J. Kasprzak; Polymer Degradation and Stability, **42**, 181 (1993).

[10] E. E. Lewis and M. A. Taylor; J. American Society, **69**. 1968 (1947).

[11] B.O. Loopstra and P. Boldrini; Acta Cryst **21**, 158 (1966).

[12] S. T. Triantafyllou, P. C. Christidis and Ch. B. Lioutas; J. Solid State Chem. **113**, 479 (1997).

[13] R. J. Cava, A. Santoro, D. W. Murphy, S. M. Zahurak and R. S. Roth; J. Solid State Chem. **50**, 121 (1983).

[14] A. Magneli; Acta Chem. Scand. 5, 670 (1951).

[15] R. D. Shannon and C. T. Prewitt; Acta Cryst. **B25**, 925 (1969).

[16] B. W. Brown and E. Banks; J. Am. Chem. Soc. **76**, 963 (1954).

[17] J. M. Berak and M. J. Sienko; J. Solid State Chem. **2**, 109 (1970).

[18] N. R. Werthamer, E. Helfand, and P. C. Hohenberg; Phys. Rev. **147**, 295 (1966).

[19] A. M. Clogston; Phys. Rev. Lett. **9**, 266 (1962).

[20] F. C. Zumsteg; Phys. Rev. B, **14**, 1406 (1976).

[21] H. R. Shanks, P. H. Slides and G. C. Danielson, Non-Stoichiometric Compounds, Advance in Chemistry Series, Vol. 39 (American Chemical Society, 1963), p. 237; A. S. Ribnick, B. Post and E. Banks, ibid. 39 (1963) 246.

[22] Q. Zhong, J. R. Dahn, and K. Colbow; Phys. Rev. B, **46**, 2554 (1992).

[23] Howard R. Shanks; Solid State Communications **15**, 753-756 (1974).

[24] A. R. Sweedler, C. J. Raub and B. T. Matthias; Phys. Letters, **15**, 108 (1965).

[25] J. B. Goodenough; Bull. Soc. Chim. France, 1200 (1965).

Figures

**Fig. 1** Observed (open circles), calculated (solid line) and difference (lower solid line) x-ray diffraction patterns of $WO_{3-x}F_x$ for $x = 0.11$ (a) and $x = 0.41$ (b) measured at room temperature. Tick marks indicate the position of allowed reflections. (c) Crystal structure of the cubic phase of $WO_{3-x}F_x$. (d) Schematic figure showing the relationship between the monoclinic, tetragonal, and cubic cells of $WO_{3-x}F_x$.

**Fig. 2** (a) Electrical resistivity $\rho(T)$ for $WO_{3-x}F_x$ with various fluorine contents. $x^*$ indicates a fluorine content estimated from lattice constant measurements; others determined analytically. (b) Temperature dependent Hall coefficient $R_H$ of $WO_{3-x}F_x$ ($x = 0.45$). The inset shows the field dependence of Hall resistivity $\rho_{xy}$ at each temperature; the solid line represents the least squares fit.

**Fig. 3** (a) The superconducting transitions of $WO_{3-x}F_x$ for various fluorine contents. $x^*$ is an estimated fluorine content from the lattice constant. The inset shows the temperature dependence of resistivity at high temperatures. (b) Electrical resistivity $\rho(T)$ of $WO_{3-x}F_x$ ($x = 0.41$) under magnetic field, with arrows marking the $H_{c2}(T)$ values. Inset shows $H_{c2}$-$T$ phase diagram for $WO_{3-x}F_x$ ($x = 0.41$). Dotted line represents linear fit near $T_c$. (c) Normalized electronic contribution to the specific heat $C_e/\gamma T$ around the superconducting transition without magnetic field (filled circles) and under an applied field of 1 T (open circles) in $WO_{3-x}F_x$ ($x = 0.44$). The inset displays the linear fit of $C/T(T^2)$ from 5K to 0.5 K, yielding the electronic specific heat coefficient $\gamma$.

Table 1 Nominal and actual fluorine content determined by chemical analysis for selected samples, with unit cell data.

| Composition (analyzed) | $WO_3$ [11] | $WO_{3-x}F_x$ $x = 0.11$ | $WO_{3-x}F_x$ $x = 0.41$ | $WO_{3-x}F_x$ $x = 0.45$ |
|---|---|---|---|---|
| Nominal ratio F/W | | 0.15 | 2.44 | 3.56 |
| Symmetry | Monoclinic | Tetragonal | Cubic | Cubic |
| Space group | $P2_1/n$ | $P4/nmm$ | $Pm\bar{3}m$ | $Pm\bar{3}m$ |
| $a$ (Å) | 7.297 | 5.24783(4) | 3.81219(3) | 3.81433(3) |
| $b$ (Å) | 7.539 | | | |
| $c$ (Å) | 7.688 | 3.86013(4) | | |
| $\beta$ (deg.) | 90.91 | | | |
| $V/Z$ (Å$^3$) | 52.61 | 53.153 | 55.402 | 55.495 |
| $Z$ | 8 | 2 | 1 | 1 |

Table 2 Crystallographic and Rietveld refinement data for $WO_{3-x}F_x$ ($x$ = 0.11 and 0.41).

| Composition | $WO_{3-x}F_x$ ($x = 0.11$) | | | | |
|---|---|---|---|---|---|
| Symmetry | $P4/nmm$ | | | | |
| $a$ (Å) | 5.24783(4) | | | | |
| $c$ (Å) | 3.86013(4) | | | | |
| Atom | $x$ | $y$ | $z$ | Occupancy | $U_{iso}$ |
| W  2($c$) | 1/4 | 1/4 | 0.4321(3) | 1 | 0.0034(4) |
| O 1  2($c$) | 1/4 | 1/4 | 0.986(4) | 0.963 | 0.034(3) |
| F 1  2($c$) | 1/4 | 1/4 | 0.986(4) | 0.037 | 0.034(3) |
| O 2  4($e$) | 0 | 0 | 1/2 | 0.963 | 0.034(3) |
| F 2  4($e$) | 0 | 0 | 1/2 | 0.037 | 0.034(3) |
| Rietveld agreement factors | | | | | |
| $\chi^2 = 3.41$ | $R_{Bragg} = 0.0931$ | $R_P = 0.147$ | $R_{WP} = 0.186$ | | |
| Composition | $WO_{3-x}F_x$ ($x = 0.41$) | | | | |
| Symmetry | $Pm\bar{3}m$ | | | | |
| $a$ (Å) | 3.81219(3) | | | | |
| Atom | $x$ | $y$ | $z$ | Occupancy | $U_{iso}$ |
| W  1($a$) | 0 | 0 | 0 | 1 | 0.0326(5) |
| O  3($d$) | 1/2 | 0 | 0 | 0.86 | 0.029(2) |
| F  3($d$) | 1/2 | 0 | 0 | 0.14 | 0.029(2) |
| Rietveld agreement factors | | | | | |
| $\chi^2 = 3.50$ | $R_{Bragg} = 0.0618$ | $R_P = 0.136$ | $R_{WP} = 0.178$ | | |

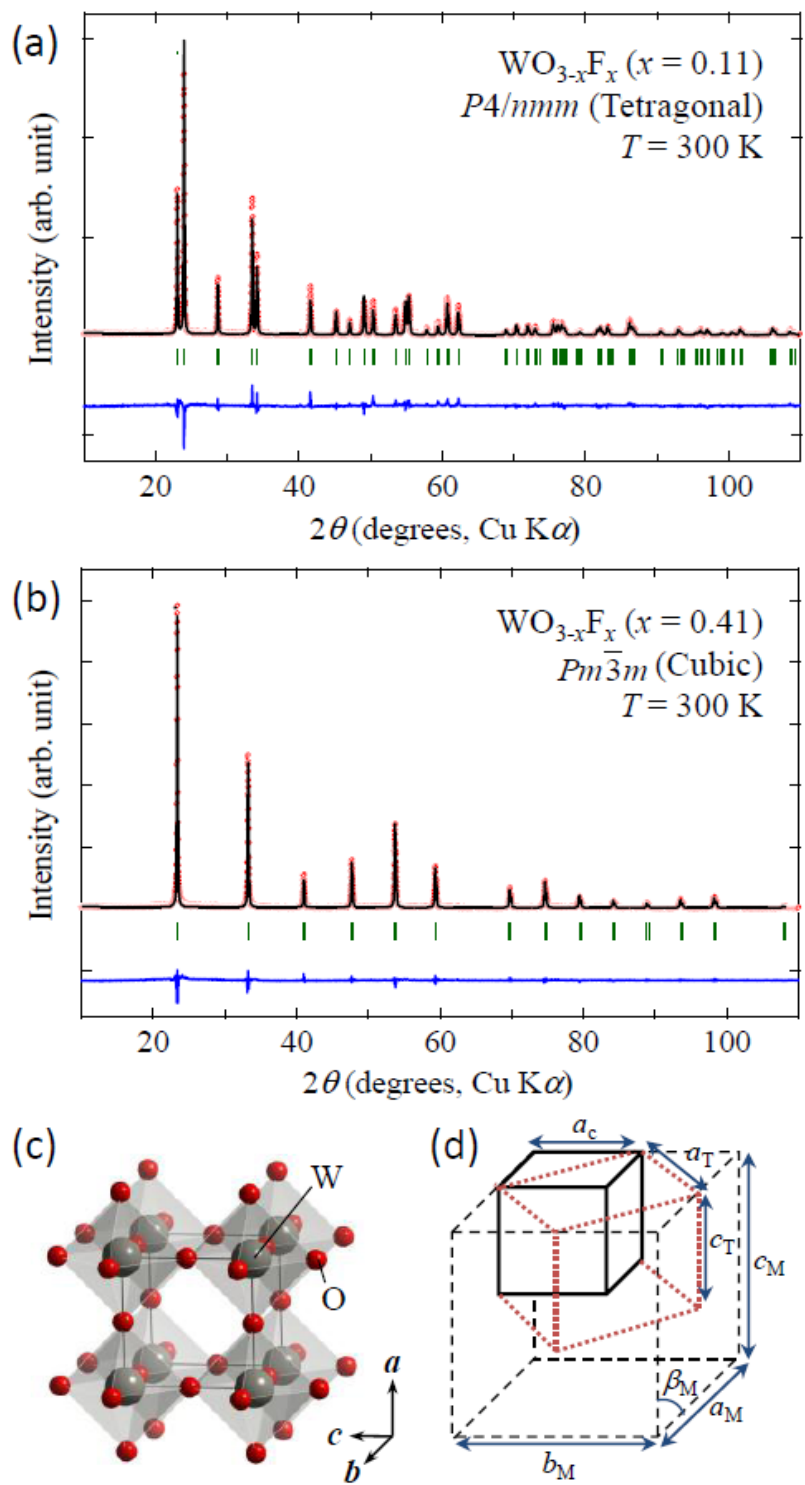

Fig. 1

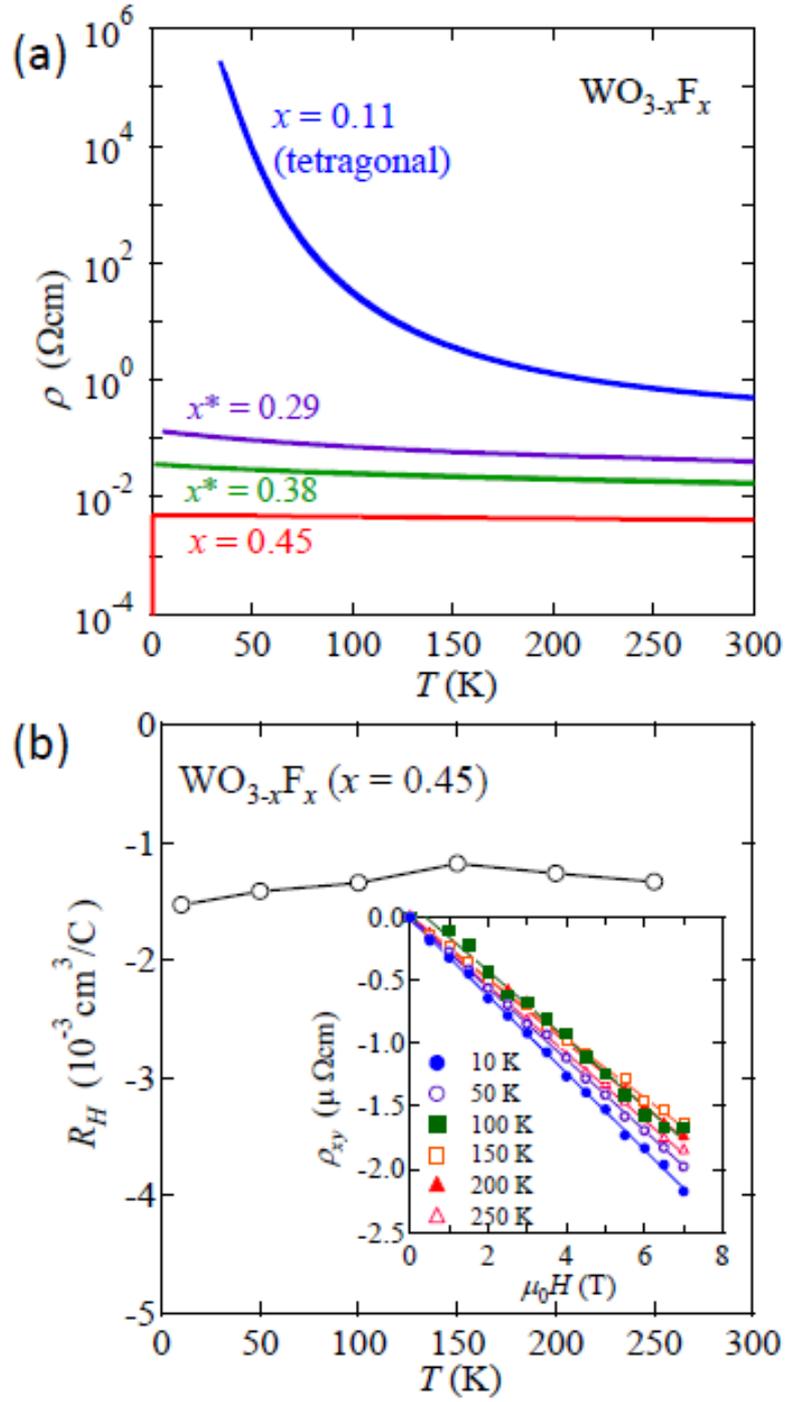

Fig. 2

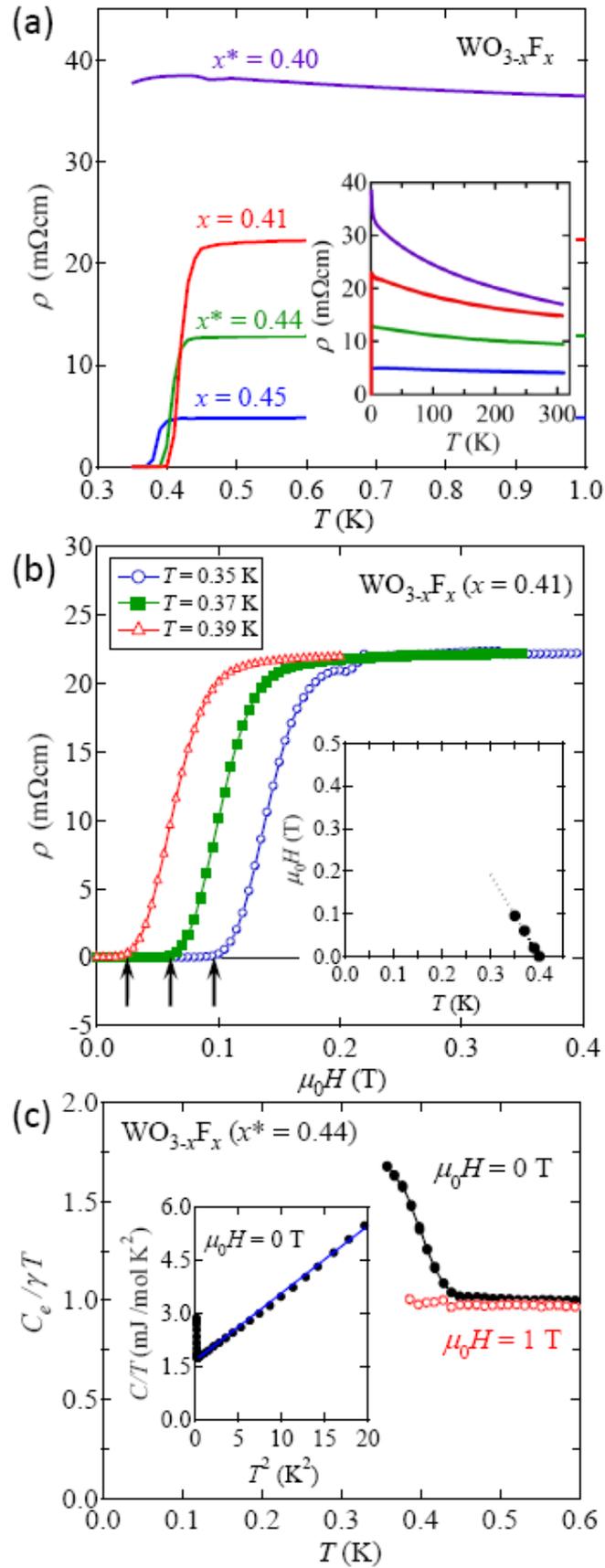

Fig. 3